\pdfoutput=1
\RequirePackage{rotating}
\documentclass[fleqn,usenatbib]{mnras}

\usepackage{newtxtext,newtxmath}

\usepackage[T1]{fontenc}

\DeclareRobustCommand{\VAN}[3]{#2}
\let\VANthebibliography\thebibliography
\def\thebibliography{\DeclareRobustCommand{\VAN}[3]{##3}\VANthebibliography}

\usepackage{graphicx}	
\usepackage{amsmath}	
\usepackage{soul}

\usepackage{url}
\usepackage{xcolor}
\usepackage[ruled,linesnumbered]{algorithm2e}
\usepackage{booktabs}

\usepackage{pdflscape}

\title[Structure of YSO Clusters]{Hierarchical Structure of YSO Clusters in the W40 and Serpens South Region: Group Extraction and Comparison with Fractal Clusters}

\author[J. Sun et al.]{
Jia Sun,$^{1, 2}$\thanks{E-mail: jiasun@pmo.ac.cn or sun.jiaaaa@gmail.com}
Robert A. Gutermuth,$^{3}$
Hongchi Wang,$^{1}$
Shuinai Zhang,$^{1, 4}$ 
Min Long$^{5}$
\\
$^{1}$Purple Mountain Observatory, Chinese Academy of Sciences,
No.10 Yuanhua Rd, Qixia District, Nanjing 210033, China\\
$^{2}$University of Chinese Academy of Sciences, No.19(A) Yuquan Rd, Shijingshan District, Beijing 100049, China\\
$^{3}$Department of Astronomy, University of Massachusetts, Amherst, MA 01003, USA \\
$^{4}$Key Laboratory of Dark Matter and Space Astronomy, Chinese Academy of Sciences, China\\
$^{5}$Department of Computer Science, Boise State University, ID 83725, USA
}

\date{Submitted to MNRAS}

\pubyear{2022}

\begin{document}
\label{firstpage}
\pagerange{\pageref{firstpage}--\pageref{lastpage}}
\maketitle

\begin{abstract}
Young stellar clusters are believed to inherit the spatial distribution like hierarchical structures of their natal molecular cloud during their formation.  
However, the change of the structures between the cloud and the young clusters is not well constrained observationally.
We select the W40 - Serpens South region ($\sim7\times9$ pc$^2$) of the Aquila Rift as a testbed and investigate hierarchical properties of spatial distribution of young stellar objects (YSOs) in this region.
We develop a minimum spanning tree (MST)-based method to group stars into several levels by successively cutting down edges longer than an algorithmically determined critical value.
A total of 832 YSOs are divided into 5 levels with 23 groups.
For describing the hierarchical properties in a controlled way, we construct a set of synthetic source distributions at various fractal dimensions, and apply the same method to explore their group characters.
By comparing the $Q$ parameter and the surface density profiles of the observed and the synthetic data, we find that the YSO observation matches spatial patterns from multi-fractal dimensions.
In the periphery region where the molecular clouds are more diffuse, the YSO structure is close to a fractal dimension of 2.0, while in the core regions the fractal dimensions are close to 1.6 and 1.4 for the W40 and the Serpens South regions, respectively.
Therefore, the YSOs may inherit the fractal pattern of the dense part of the molecular clouds, but such pattern dissipates slowly in several Myr. 
\end{abstract}




\section{Introduction} \label{sec:Intro}

When clusters of young stellar objects (YSOs) form in their natal molecular cloud, they are believed to inherit properties such as metal abundance, angular momentum, and initial spatial distributions.
For the spatial distributions, most prestellar cores are found in filamentary structures of molecular clouds, which in fact contain more than 80\% mass of dense gas \citep{Andre2011, Konyves2015, Hacar2018, Arzoumanian2019}.
These filaments compose complex hierarchical structures, and the newly born YSOs are supposed to be distributed in a similar way.
However, such structures will be disturbed by multiple factors in a few crossing time, such as gravitational interactions \citep{Ballone2020}, stellar explosions, mergers \citep{Bally2017}, and kicking \citep{Reipurth2010}. 
For this reason, there is a scenario that the hierarchical structures of YSOs at early stage should tend to diminish as they evolve toward main sequence stars \citep{Goodwin2004}.

To verify this scenario, we select  W40 - Serpens South Region in Aquila Rift as our testbed to analyze structures of YSOs inside. 
There are a number of reasons for our selection. 
First,  the sample size is sufficient. 
We identified a total of 832 YSOs in the catalog, including 15 deeply embedded sources, 135 Class I, 647 Class II, and 35 transition disk sources (Sun et al. 2022, submitted to MNRAS).
Second, these YSOs are well correlated with the molecular filaments and have differential age distributions.
For example, YSOs in the Serpens South region are generally younger than 1 Myr,  while YSOs in the W40 region are older with ages of several Myr.  
This difference can nicely distinguish the different stages of structure evolution.

A parameter called the fractal dimension can be used to describe those hierarchical structures.
When the concept of the fractal was first developed in the early 20th century, it refers to the \textit{exact} self-similarity that can be realized with infinitely recursions on computer, such as Cantor set, Peano curve, Koch snowflakes, Sierpinski triangle, and so on. 
This kind of fractal pattern can be perceived visually for nature objects such as snowflakes, broccoli, and blood vessels.
Later on, the term fractal was defined more liberally, including \textit{quasi} and \textit{statistical} self-similarity by Mandelbrot \citep{Mandelbrot1967}, whose initial intention was to use it to depict the roughness of nature objects, such as clouds, coast-lines, and clusters, which have abundant substructures but do not imitate the entire structure exactly.
Based on the new definition, the fractal applications are prevalent in astronomy.
The fractals have not only been used to describe statistical distributions from large-scale structures of the universe \citep{Scrimgeour2012} to the Saturn ring \citep{Avron1981}, but also to explain accelerated speed of dynamic processes such as mixing and cooling of winds and shocks in clouds \citep{Sutherland2003, Banda-barragan2019}, and dust aggregating in planetesimals \citep{Blum2008}.

Fractal models have been used to characterize substructures in molecular gas clouds.
To explain the power-law distributions of size and mass in molecular clouds, \citet{Elmegreen1996} interpreted the fractal dimension $D=2.3$ with a turbulent origin.
Later on, more studies on the degree of fractal geometry of molecular clouds show the fractal dimension varies in a range between 2.2 and 2.8 \citep[e.g.,][]{Dickman1990, Stutzki1998}.
However, most studies are based on the perimeter-area method and deduce a projection of fractal dimensions, which does not have a clear correlation to the 3D fractal dimension though \citep{Sanchez2007}.
Thus, the fractal dimensions for those dense molecular filaments that give birth to YSOs might be different, especially when they get affected by feedback from forming stars.

Fractal models have also been used to characterize the clustering of star-forming sites in nearby galaxies.
It is very likely that young clusters are of smaller fractal dimensions than those for the clouds.
\cite{FeitzingerGalinski1987} studied the patchiness of 7644 H {\sc ii} regions as tracers of star-forming sites in 19 spiral galaxies, and obtained a roughly constant fractal dimension of $1.68\pm0.31$ for all galaxies.
However, in another sample of 10 galaxies, \cite{Elmegreen2001} got the fractal dimension of star formation of $\sim$2.3, which coincides with the value for the molecular clouds \citep{Elmegreen1996}.
In the solar neighborhood, \cite{delaFuenteMarcos2006} suggested multi-fractal structures in the young open clusters.
It means, for clusters younger than 20 Myr, the fractal dimension varies from 0.61 to 1.65, while for clusters with ages in 20--60 Myr, it varies from 1.13 to 1.85. 
The gas structure that formed one generation of clusters is somewhat different than for the next.

In this work we investigate hierarchical structure with YSOs from one molecular cloud, tracing substructure among adjacent over-densities down to very small size scales ($\sim$0.1 pc).  
However, currently there are no standard tools to estimate the fractal dimensions directly from observed structures in a 2D map.
We develop an extended grouping method based on the core extraction method described by \citet{Gutermuth2009} (hereafter \textbf{G09}), and extract hierarchical structures from the YSOs and compare them to fractal model clusters.

There are multiple ways to subdivide groups in a cluster, such as constructing stars' surface density \citep[e.g.][]{Panwar2019}, and K-nearest neighbor analysis which aim to separate clusters with nearest mean distance to cluster centers \citep[e.g.][]{Pasztor1992,Joncour2018}.
\cite{Megeath2016} identified clusters using a ``friend of a friend'' technique, which combines the surface density with N nearest neighbors.
Some methods are parametric, e.g. \cite{Kuhn2014} performed finite mixture models of isothermal ellipsoids to identify young subclusters.
The cluster's features can also be interpreted with Two Point Correlation Function (TPCF) \citep[e.g.][]{Grasha2015,Zhangmiaomiao2019}, which describes neighbors distribution of a cluster within a certain radius and annulus. 
People have also used dendrogram to study the the structure trees of the clusters \citep[e.g.][]{Joncour2019}, and $\Delta$-variance wavelet transform technique
to study their self-similarity \citep[e.g.][]{Gouliermis2014}.
The G09 method that we use is graph-based algorithm, and it has good tests on fitting cumulative distribution function (CDF) of minimum spanning tree (MST) to extract cluster cores out from diffuse peripheral stars.

The paper is organized as follows. 
Section \ref{sec:Data} presents observed structures of our identified YSOs in the W40 and Serpens South region, and synthetic models of clusters with different fractal dimensions generated by the \texttt{McLuster} explanation and reference routines.
We develop the grouping method for the hierarchical structures in Section \ref{chap:Core}, and apply it to both observed and synthetic data of YSOs in Section \ref{sec:group_analysis}.
In Section \ref{sec:Fd}, we compare the observed YSOs data and the synthetic models in terms of $Q$ parameter, surface density profiles, and discuss the evolving fractal structures of YSOs.
Summary is provided in Section \ref{sec:sum}.

\section{Observed YSO and Synthetic Clusters} \label{sec:Data}

\subsection{The W40 - Serpens South YSO Cluster}
Sun et al. 2022 (submitted to MNRAS) presented a comprehensive catalog containing 910,275 point sources in the  W40 - Serpens South Region.
The catalog is constructed using multiple surveys which are complementary to each other:  near-infrared (NIR) WIRCam observations with the CFHT telescope, 2MASS and UKIDSS data for the NIR sources, and the Spitzer data for the mid-infrared (MIR) emissions.
In this comprehensive catalog, 832 YSOs are identified, including 15 deeply embedded sources, 135 Class I, 647 Class II, and 35 transition disk sources. 

\begin{figure*}
\centering
\includegraphics[width=0.9\textwidth]{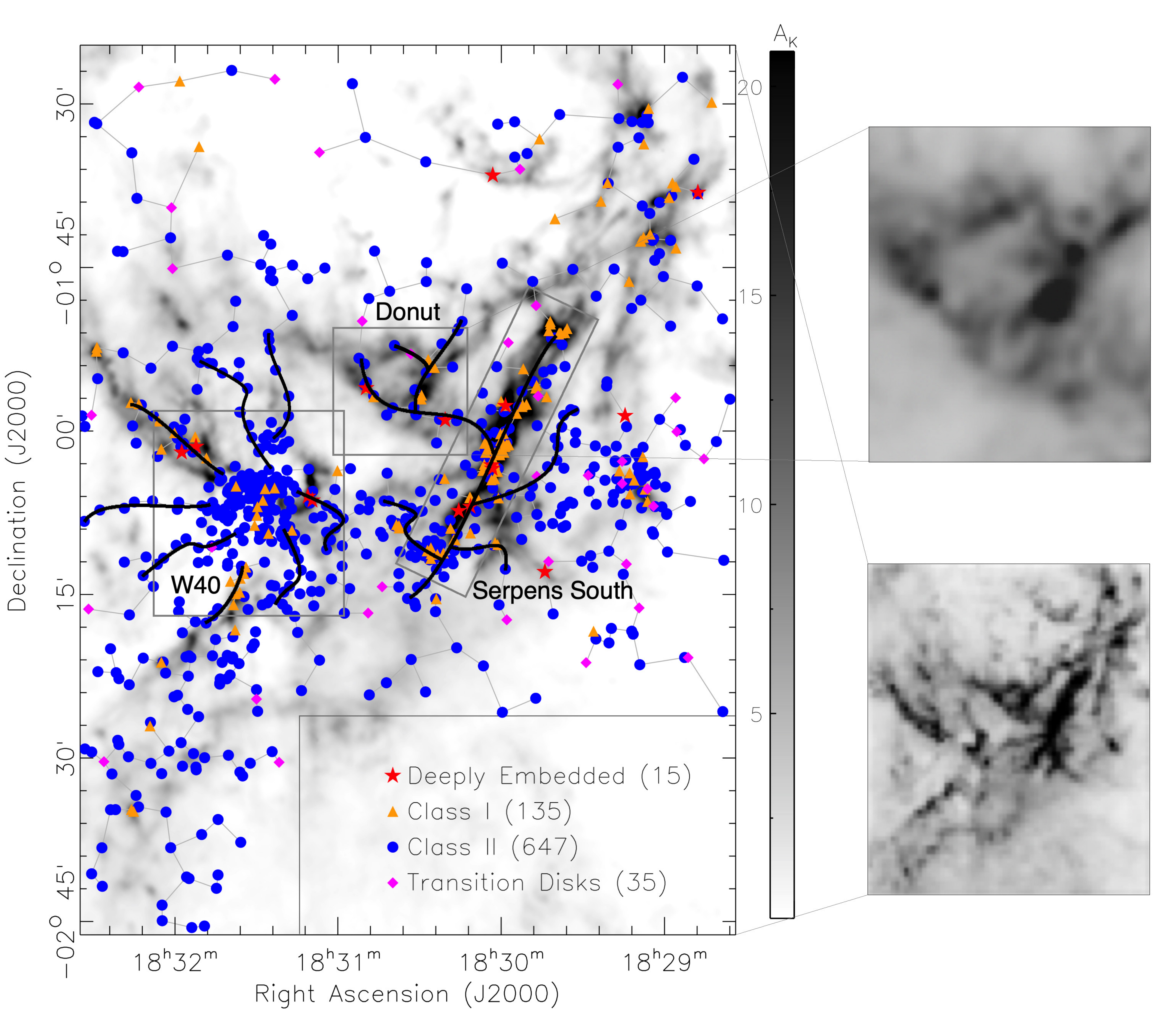}
\caption{
The left panel shows the distribution of the 832 classified YSOs in the W40 - Serpens South region, which are overlaid on the molecular clouds derived from Herschel hydrogen column density map \citep{Andre2010}.
All the YSOs are connected by grey edges to form a MST, while the main branches of the W40 and the Serpens South clusters are sketched by the bold black curves.
The YSOs in the W40 region show a spider-like structure and that in the Serpens South region show a gecko-like structure. 
The right panel shows two molecular cloud maps of the ``Donut'' Cloud and the entire region, which have a similarity in shapes at the same pixel resolution.
}
\label{fig:YSOsMST_demo}
\end{figure*}

Figure \ref{fig:YSOsMST_demo} shows the distributions of YSOs, overlaid on the map of molecular clouds from Herschel Data \citep{Andre2010}.
Along the central ridge of the Serpens South region, massive YSOs are found to have ages mostly less than $\sim$ 0.46--0.72 Myr \citep{Plunkett2018}, while the chemical evolution model suggests the star forming age as about less than 0.3 Myr \citep{Friesen2013}.
Consistently, the high percentages ($\sim$44\%) of Class I YSOs identified in the Serpens South region are representing a younger cluster age of about less than 1 Myr.
In W40, YSOs are dominated by Class II, which suggests that the age of the W40 cluster is in the order of a few Myrs.
Based on the existence of OB stars in the central region of W40, consistently, \citet{Shuping2012} gives an upper limit of the cluster age of about 7 Myrs.
It is likely that the molecular gas has been exhausted by the star formation or expelled by these massive OB stars in the W40 region.

We should notice that our YSO category does not have a complete inventory of “YSOs”.
In the infrared, diskless members are confused with and dominated by unrelated field stars along the line of sight, thus we exclude them to keep close to a member-dominated source sample.
Nevertheless, the goal of this work is to investigate the young stellar clusters that are just given birth, whose structure may inherit that of the dense part of the molecular clouds but could also have changed a bit.
Our category that covers the stages of Class I--II within several Myrs is a relatively representative sample.
The duration at Class III stage is generally longer than that at Class II stage, and thus the structure of Class III YSOs can be studied separately.

More than 800 YSOs in our category should be capable of revealing the basic fractal properties, such as the Q parameter in Section \ref{chap:qparameter}.
\cite{Bastian2009} has qualified the extinction effects on Q, which is not significant when the number of stars is relatively large ($\sim$1000).
In addition, the completeness of our Class I--II YSOs is relatively high, estimated by the proportional correlation of front extinction of YSOs and column density of molecular clouds at their positions (Sun et al. 2022, submitted to MNRAS).
This ensures the effectiveness of the comparison with the synthetic clusters.

\subsection{Synthetic clusters generated using McLuster}\label{chap:mcluster}

To provide a quantitative way to analyze the above observed cluster structure, we also need to generate a set of synthetic clusters under various given fractal dimensions by using the \texttt{McLuster} procedure.
This code is designed to study mass segregation and fractal substructure in star clusters \citep{Kupper2011} by taking the following steps.
First, it initializes a box with a fixed size, and inputs a star with a small random offset to the center of the box.
Second, the box is split into eight half-sized boxes.
The probability that a sub-box gets a new star is then $2^{D-3}$, where D is the fractal dimension. 
Third, the above step is repeated till $128.0\times8^{log(N)/log(8)}$ stars are generated, and then $N$ stars within a certain distance are randomly taken out. 
Given that the observed structure of YSOs are 2D, we add a fourth step, which projects the 3D structure generated in the previous steps along the x/y/z axis onto a 2D map.
This step is necessary and enables us to make direct comparisons between synthetic structures and observed 2D structures on the sky plane.

In total, we generate a set of synthetic clusters with fractal dimensions varying from 1.0 to 2.0 with gradient of 0.1, and an additional fractal dimension of 2.7.
For each fractal dimension we initialize 170 3D synthetic clusters with $N=800$, and project them along the x/y/z axis to obtain 510 2D cluster maps. 

\section{Hierarchical Grouping Method}\label{chap:Core}

\subsection{CDF of MST}\label{chap:CDFofMST}

We use a MST-based method to represent the complex spatial distribution of stars, for example, the filamentary structures.
The MST, or ``minimum-weight spanning tree'' is a concept in graph theory, representing an acyclic subset of edges ``spans'' in a connected, weighted, undirected graph.
It connects all vertices to form a tree structure, but pushes the total edge weights to be the minimum.
The MST is frequently used to separate close stellar clusters \citep[e.g.][]{Battinelli1991, Hetem1993, Cartwright2004, Gutermuth2009, Kirk2014}. 

Fig. \ref{fig:YSOsMST_demo} shows the MST for the  W40 - Serpens South YSOs.
Vertices depict YSOs, edges mean line segments between YSOs, and weights are the lengths of those segments.
It can be seen that the W40 cluster and the Serpens South clusters are well separated and loosely connected by a single edge.
In the W40 region, there are eight MST main branches spreading to all directions, in the shape of a spider.
In the Serpens South region, four MST main branches join to a filamentary body perpendicularly, composing a gecko-like shape.
A small group of YSOs lying north east of the Serpens South has a donut-shape distribution, matching well the outline of its molecular cloud, and thus we name it as “Donut” now that it is relatively independent. 
We show in Fig. \ref{fig:YSOsMST_demo} the similarity between the “Donut” cloud and the cloud of the entire region which happens to have a hollow structure.

In the complete MST, some YSOs are more concentrated, while some are sparse and loosely linked.
To determine proper hierarchical levels of YSOs in the entire target region, we develop an extended grouping method using the CDF of MST.
CDF is a percentage of the number of edges which are shorter than a critical value compared to the total number of edges, and thus reflects the distribution possibility of the edge length of MST.
The concentrated clusters tend to have shorter edge lengths, while those loosely linked have longer lengths.
Therefore, the two kinds of structures are distinguishable in a CDF.

\subsection{An ideal CDF of MST with randomly and evenly distributed vertices}\label{chap:random}
Observed stellar distributions that are consistent with the statistical variations of a naturally uniform distribution should not be subdivided.  
A cluster of randomly and evenly distributed vertices can be used as an ideal reference to demonstrate this.
Accordingly, we perform numerous Monte Carlo trials, scattering various $N$ number of points in a fixed box with an area $A$. 
For each trial, we construct its MST and compute the associated CDF.
Our analysis of these trials reveal a reasonable reproduction of the ideal CDF (see \cite{Bronshtein1998}) by Equation \ref{eq:cdf}:
\begin{equation}\label{eq:cdf}
	CDF_\mathrm{uniform} = C \cdot \Sigma^{\frac{7}{8}} \int_{x=0}^{\infty} x^{\frac{3}{4}} e^{-\frac{(x-\mu)^2}{2\mu^2}}dx,
\end{equation}
where, $x$ is the MST edge length, $\Sigma = \frac{N}{A}$ is the mean surface density, and $\mu = \frac{1}{ \sqrt{2\pi \Sigma}}$ is approximately the typical MST edge length. 
$C=\frac{1}{0.49}$  is a constant that get this integral normalized.

\begin{figure*}
\centering
\includegraphics[width=0.95\textwidth]{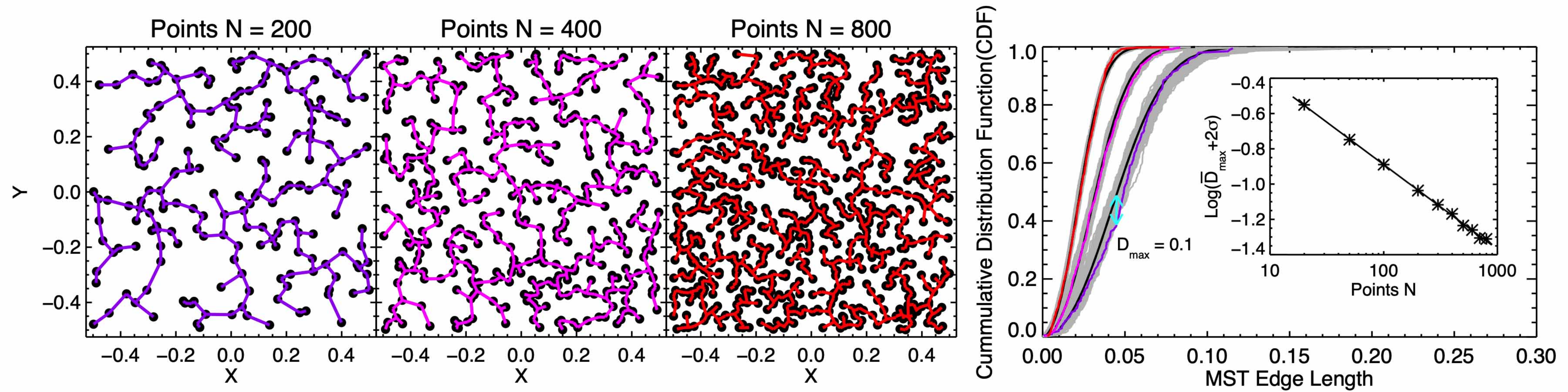}
\caption{
Left panels: three MST samples with randomly distributed points. 
Right: the CDF curves (red, magenta, and purple are corresponding to $N=$200, 400, and 800 points, respectively) and the ideal CDF curves (black) due to Eqn. \ref{eq:cdf}. Overlaid grey CDF curves are from the other 399 data sets with the same $N$ points. The fitting of $N$ and the tolerance (i.e. $\overline{D}_{\rm max}+2\sigma$) is shown in the inlet figure.}
\label{fig:random_cdf}
\end{figure*}

Fig. \ref{fig:random_cdf} shows three MST samples of randomly distributed points with $N=$200, 400, and 800. 
For a larger $N$, the $\Sigma$ increases, and the edges are typically shorter as to show a steeper slope in its CDF curve.
We compare their CDF curves with the ideal CDF from Eqn. \ref{eq:cdf}, and find that the difference between them always exists.
Thus, for each trial, we record the maximum distance $D_{max}$ between the two curves to represent for the difference, which is $\sim$0.1 for the $N=$200 case for example.
In order to determine a tolerance for statistical variance, we calculate $D_{max}$ for 400 sets of random points with $N$ from 100 to 800, in a step of 100.
From $N =$ 100 to 800, the mean $\overline{D}_{max}$ varies from 0.167 to 0.032, and its $\sigma$ varies from 0.058 to 0.0084.
We set the $\overline{D}_{max}+2\sigma$ as the tolerance that includes 97.85\% cases, and find a linear correlation between ${\rm log(}tolerance{\rm)}$ and ${\rm log(}N{\rm)}$, as shown in the inlet figure of Fig. \ref{fig:random_cdf}.
Therefore, we determine that a group with $D_{max}$ less than the interpolated tolerance value $10^{(-0.48{\rm log}N+0.075)} $ is consistent with a random distribution and will not be further divided.

\subsection{Grouping method and the illustration}\label{chap:GroupingMethod}

\begin{figure*}
\centering
\includegraphics[width=0.95\textwidth]{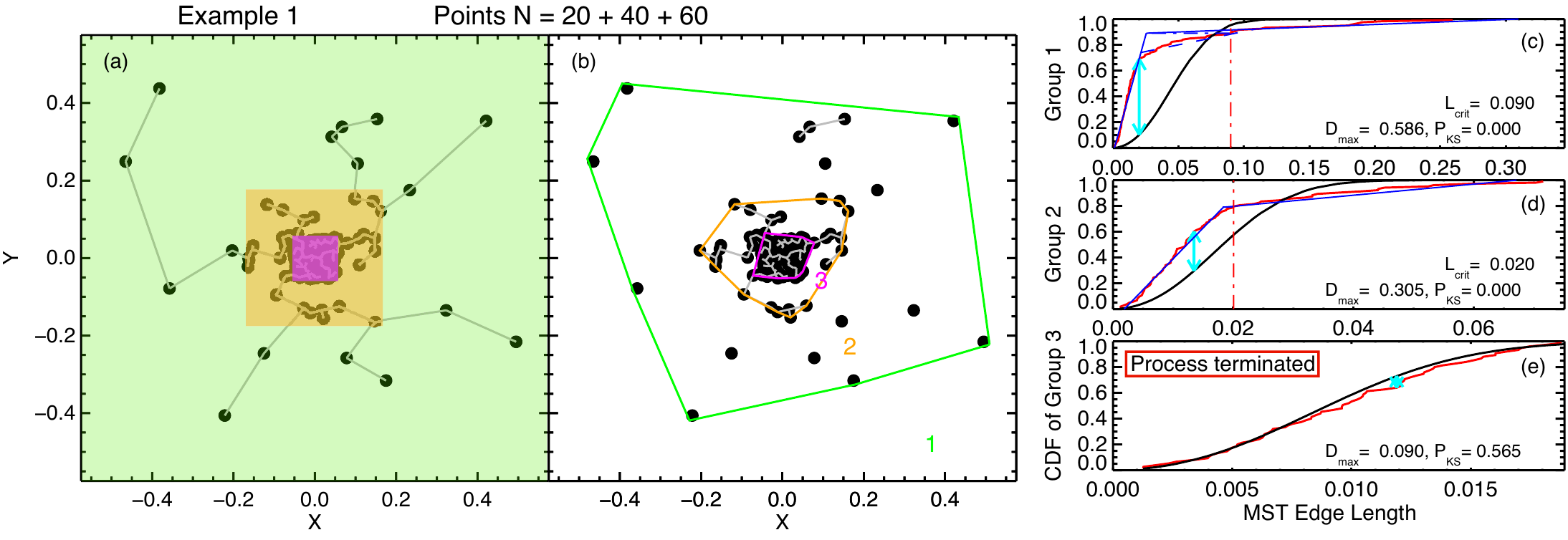}
\includegraphics[width=0.95\textwidth]{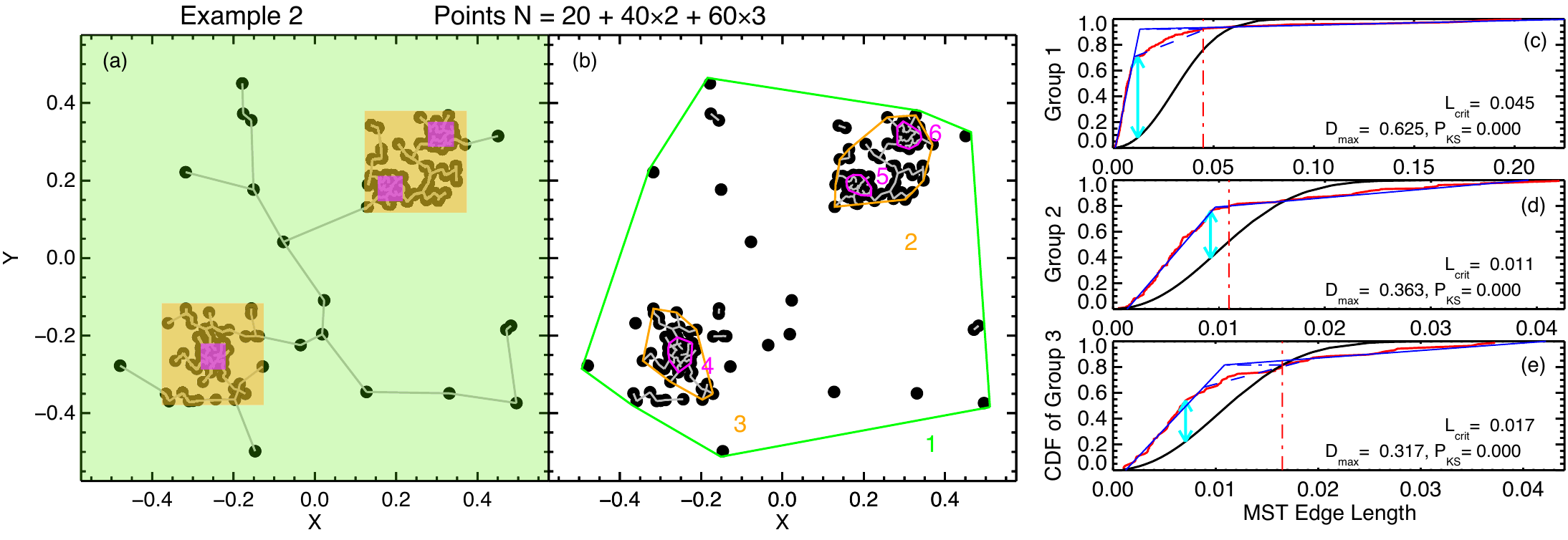}
\caption{
Two examples are used for the demonstration of our grouping method for searching the hierarchical structures. \newline
`Example 1' is expressed in the upper panel: (a) shows how the artificial cluster is constructed, where 20, 40, and 60 points are randomly distributed in the light green box, the orange box, and the central purple box, respectively. All the points are connected by grey edges to form a MST. (b) shows the same points and the results after applying our grouping method. The `Group 1' containing all points is outlined by the green convex hull, which is at the level 1. The `Group 2' at the level 2 and the `Group 3' at the level 3 are outlined by the orange and purple convex hulls,  respectively. All the edges longer than the critical length at the level 1 are cut for an illustration of how the `Group 2' is recognized. (c), (d), and (e) are the CDF curves for the `Groups 1, 2, and 3' (red), compared with their ideal CDF curves (black). In (c) and (d), the $D_{\rm max}$ values (cyan) are large and thus their CDF curves are fitted by a polyline to determine the critical edge length. In (e) however, the $D_{\rm max}$ is smaller than the tolerance value and the K-S test is not passed. As a consequence, the grouping process is terminated at the `Group 3'. \newline
`Example 2' is expressed in the lower panel: (a)--(e) are similar to the above. But in (a), the point distribution is more complex, where 20 points, two sets of 40 points, and three sets of 60 points are randomly distributed in a light green box,  two orange boxes, and three purple boxes, respectively. In (b), 6 groups are recognized at three hierarchical levels. All the edges longer than the critical length at the level 1 are cut as well. (c), (d), and (e) are the CDF curves for the `Groups 1, 2, and 3' (magenta). The grouping process is terminated at the `Group 4, 5, and 6', though not presented here.
}
\label{fig:eg_cdf}
\end{figure*}

With the ideal CDF curves for random distributions and their tolerance values, we can distinguish different patterns and hierarchical structures of spatial distributions.
Our grouping method is based on the core extraction method in G09, which can tell the pattern of cores from more discrete sources in a CDF curve.
G09 did a survey of 36 nearby YSO clusters with numbers $N$ varying from 3 to 235, among which 8 clusters have two cores, 22 clusters have one core, and 6 clusters have no core.
In the W40 - Serpens South region, however, there are $N=832$ YSOs, which indicates the possibility of substructures in the ``cores'' extracted by the G09 method.
Therefore, our grouping method will perform the similar procedure to the ``cores'' again and recursively, until the CDF differences are within the tolerance, to reveal the possible hierarchical structures.

We give an illustration here for our grouping method based on two simple artificial clusters.
For the first example cluster, we add points in the following way.
First, 20 points are randomly distributed in the entire light green region.
Second, 40 points are randomly scattered in the central orange box.
Last, 60 additional points are randomly put in the central smaller purple box, as shown in Fig. \ref{fig:eg_cdf}a of `Example 1'.
Consequently, there are 3 sets of points and 3 spatial patterns with differential surface densities. 
The second example cluster has a slightly complex distribution, where 20 points, two sets of 40 points, and three sets of 60 points are randomly distributed in a light green box,  two orange boxes, and three purple boxes, respectively (Fig. \ref{fig:eg_cdf}a of `Example 2'.).
Finally, there are 6 sets of points, but still 3 spatial patterns.

In `Example 1',  the points are all connected to form a MST, and bounded by a convex hull as `Group 1' (Fig. \ref{fig:eg_cdf}b). 
The convex hull is the smallest convex polygon that contains all the points.
The `Group 1' CDF clearly deviates from the CDF$_\mathrm{Random}$ (Fig. \ref{fig:eg_cdf}c), as it contains a large amount of short edges, indicating more complex substructures.
A group that can be divided into sub-groups is called as a parent group, when the following conditions are satisfied:
1, the $D_{\rm max}$ difference between its CDF and the ideal CDF$_\mathrm{Random}$ with the same $N$ is greater than the tolerance value.
2, it should pass the Kolmogorov-Smirnov (K-S) test with a possibility $p<0.05$.
3, a parent group is required to have a number of points $N_{\rm parent}>20$, while a child group is required $N_{\rm child}>10$.
Child groups can have their group properties meaningfully examined with more than 10 memberships, which is a commonly used cutoff value \citep[e.g.][]{Kirk2011,Gutermuth2009,Getman2018}. 
Meanwhile parent groups are expected to have 20 memberships so that they can contain some practical periphery stars besides their child groups.
A larger cutoff enlarges the S/N of points in the groups, which is stabler when membership changes.
The lower it goes, the more tradeoff we have to make between completeness and robustness of the groups. 
In the current case, the `Group 1'  is a parent group,  and may have one or several child groups.

We fit the CDF curve of the `Group 1' by a two or three-segment polyline, and get an intersection point of the first and the last segments.
From the exact edge length of this point, to the length at which the curve reaches the cumulative distribution value of this intersection point, we obtain a range of transition edge lengths, altering from lower surface density points to higher.
In G09, the former, i.e. the exact length of the point, is taken as the critical value $L_{\rm crit}$, because the fitting is performed only once to find dense cluster cores. 
When a similar method is applied to large-N source distributions with a wider density dynamic range, lower density structure is often ignored and only the denser groups are isolated \citep[e.g.,][]{Koenig2008}.
Here we adopt the latter to be $L_{\rm crit}$, which is larger and retains more substructures in a complex hierarchical cluster (Fig. \ref{fig:eg_cdf}c and d).

All edges with longer length are cut, as shown in Fig. \ref{fig:eg_cdf}b.
The points that are still connected by edges are considered to be child groups if  $N>10$.
The other points that are detached and not assigned to any group are in the first hierarchical level, which is called level 1.
The `Group 1' has only one child group, which is called as the `Group 2' ($N=104$).
At this stage, we find a deeper level of the entire structure and call it the level 2, and the `Group 2' is at this level 2.
The `Group 2' satisfies the three conditions and turns out to be a parent group too.
We apply the same procedure recursively, find the critical length value $L_{\rm crit}$ (Fig. \ref{fig:eg_cdf}d), and get the child `Group 3' ($N=76$) from the `Group 2'.
The `Group 3' is at level 3.
However, $D_{\rm max}$ of the `Group 3' is less than the tolerance value and the K-S test possibility is larger than 0.05 (Fig. \ref{fig:eg_cdf}e), and therefore the process is terminated.
As last, 3 hierarchical levels are identified correctly, and the artificial groups are reasonably outlined by their own convex hulls.


Based on the same grouping method, 6 groups are identified in `Example 2', but the hierarchical levels are also three.
In summary, the grouping method can successfully detect and distinguish various hierarchical structures from MSTs.

\section{Grouping analysis}\label{sec:group_analysis}

\subsection{ W40 - Serpens South YSO Cluster Grouping}\label{chap:YSOgrouping}

\begin{figure*}
\centering
\includegraphics[width=0.367\textwidth]{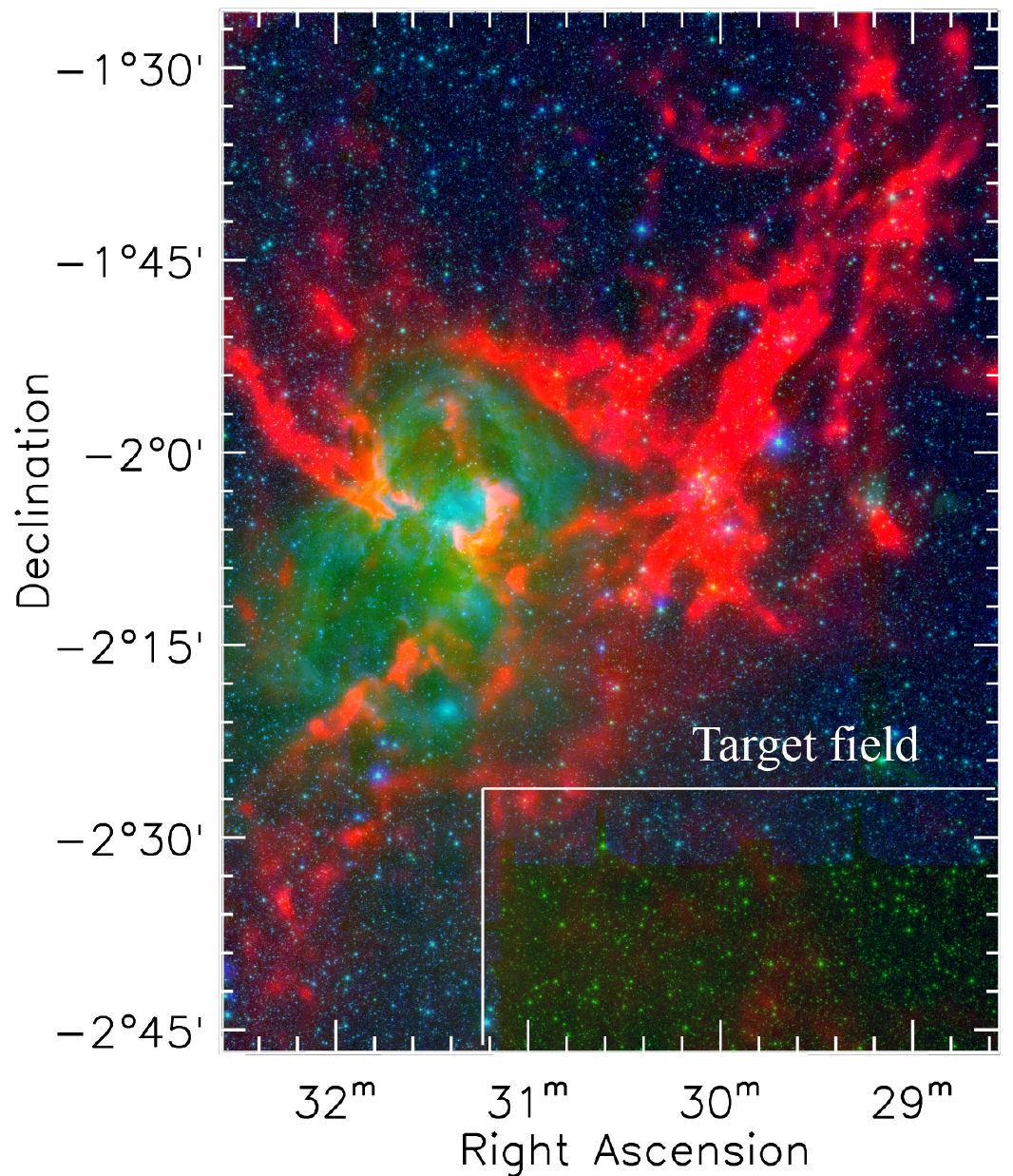}
\includegraphics[width=0.623\textwidth]{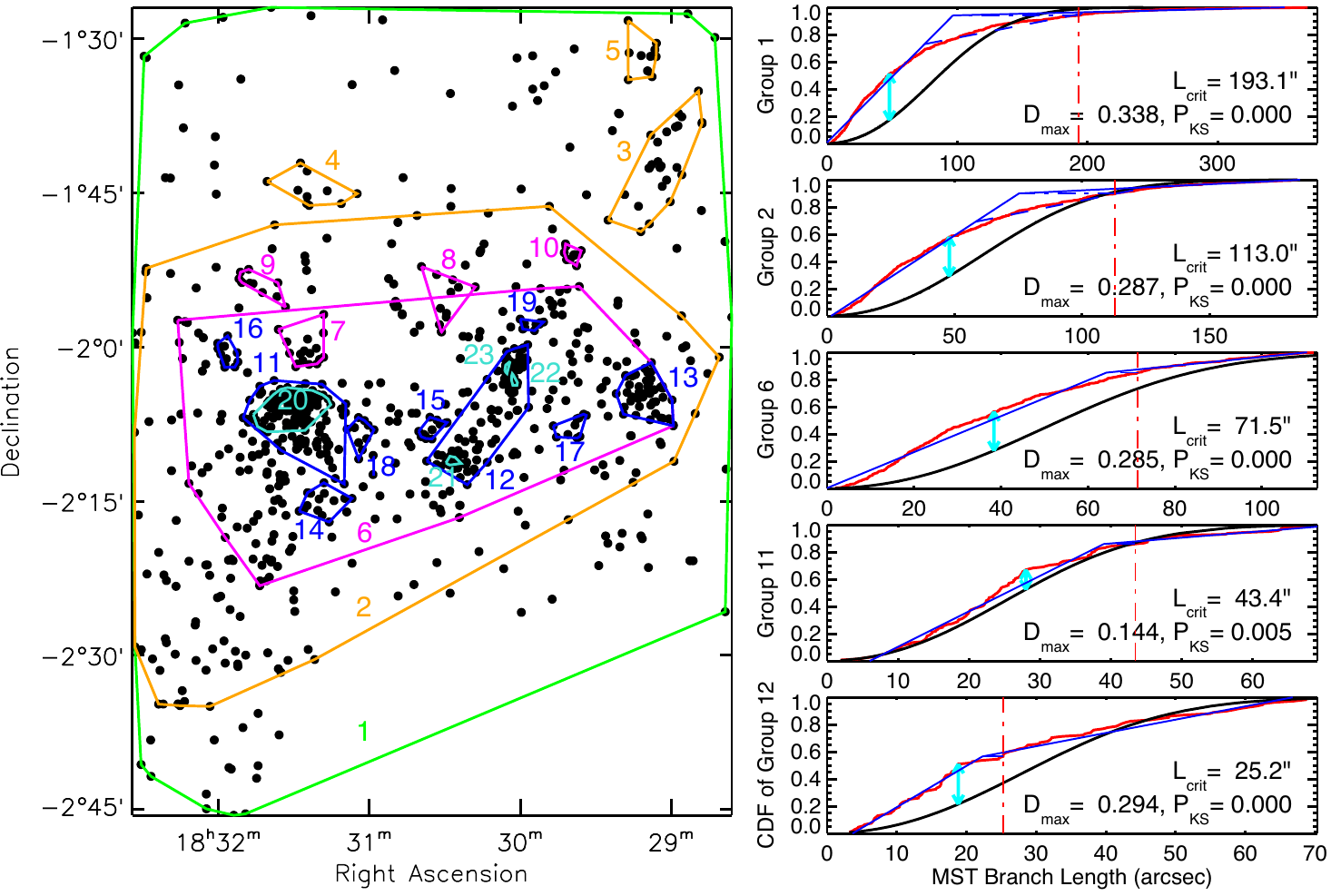}
\caption{ 
Left panel: composite IR image of W40 and the Serpens South. 
The Target Field was observed by WIRCam in Ks band (blue) and is combined with Spitzer 4.5 $\rm \mu$m (green) and Herschel hydrogen column density (red).
Middle panel: the groups and the hierarchical structures of the YSO cluster in W40 - Serpens South region. 
The 5 level structures from outside to inside are marked with different colors.
Labels from 1 to 23 represent groups at different levels, as referred in Table \ref{tab:grpinfo}.
Right panels: the CDF curves of important groups, the same as in Fig. \ref{fig:eg_cdf}.}
\label{fig:YSOs_Grouping}
\end{figure*}

We apply the above grouping method to the YSOs in the W40 - Serpens South region and obtain a 5-level structure of YSOs shown in Fig. \ref{fig:YSOs_Grouping}.
At levels 1--3, each level has only one parent, and their critical values of edge length are 193.1$''$, 113.0$''$, and 71.5$''$ (which are 0.41 pc, 0.24 pc, and 0.15 pc at the distance of 436 pc).
At level 4, however, there are two parent groups, corresponding to the core regions in the W40 and in the Serpens South, and their critical lengths are 43.4$''$ and 25.3$''$ (which are 0.092 pc and 0.053 pc).
The critical lengths are decreasing by a mean factor of about 1.8.
We denote a total of 23 groups across all 5 levels using group ID from 1 to 23, as shown in Table \ref{tab:grpinfo} for the following discussion.

The groups 2--5 are at level 2. 
The northern child groups 3--5 that show local concentration of sources are detached from the central region (i.e., the group 2).
Other sparse YSOs may be still connected by edges but have the member stars less than 10, and thus are not qualified to be child groups.

The groups 6--10 are at the level 3, all cut off from the parent group 2.
Among them, the groups 7--10 are not far away from the group 6, and their convex hulls are even overlapped.
The group 8 is at the right side of the Donut cloud, while the YSOs at the left side are detached.
Nevertheless, this Donut cloud has the densest molecular cloud around the group 8.
The group 10 is inside the Serpens South region, though it seems slightly independent.
The other sparse YSOs at the level 3 have higher surface density compared to those sparse ones at the level 2, and are at regions with denser molecular clouds.

The groups 11--19 are at level 4, all cut off from the parent group 6.
The W40 cluster (the group 11) and the Serpens South cluster (the group 12) are well separated, where the former has a rounder convex hull while the latter has an elongated shape.
The groups 13--19 are likely corresponding to other stellar clusters.

The groups 20--23 are at level 5, in which the first one is inside the parent group 11, and the other three are cut from the group 12.
They are the densest cluster cores inside the star forming regions.
The statistics of some properties of all these groups are listed in Table \ref{tab:grpinfo}.

\begin{table*}
 	 \centering
	 \caption{Some statistics of the W40 - Serpens South YSO Cluster's Group Memberships.}
	 \label{tab:grpinfo}
    \footnotesize
{
	\begin{tabular}{|c|c|c|c|c|c|c|c|c|c|c|}
	\hline
GrpID & Level & Parent GrpID & Ra & Dec & R$_{h}$(Arcmin$|$pc) & N (star) & $\Sigma$(*/Deg$^2$)  & Q param & Class I/II\\
	\hline
\textcolor{green}{ 1}& 1&NaN& 18:30:40.912& -2:2:44.584&  36.27 $|$  4.60&  832&   724.79& 0.740& 0.220\\
	\hline
\textcolor{orange}{ 2}& 2& 1& 18:30:56.338& -2:7:36.044&  25.16 $|$  3.19&  696&  1260.10& 0.632& 0.228\\
\textcolor{orange}{ 3}& 2& 1& 18:29:4.818& -1:42:36.331&   5.15 $|$  0.65&   26&  1125.38& 0.690& 0.625\\
\textcolor{orange}{ 4}& 2& 1& 18:31:22.977& -1:44:23.320&   3.51 $|$  0.45&   10&   929.08& 0.788& 0.000\\
\textcolor{orange}{ 5}& 2& 1& 18:29:12.275& -1:31:27.388&   2.78 $|$  0.35&   10&  1487.23& 0.881& 0.250\\
	\hline
\textcolor{magenta}{ 6}& 3& 2& 18:30:47.915& -2:6:36.480&  17.73 $|$  2.25&  510&  1858.10& 0.498& 0.259\\
\textcolor{magenta}{ 7}& 3& 2& 18:31:25.423& -1:59:25.224&   2.64 $|$  0.33&   18&  2963.61& 0.786& 0.000\\
\textcolor{magenta}{ 8}& 3& 2& 18:30:29.491& -1:54:56.763&   2.68 $|$  0.34&   12&  1909.60& 0.643& 0.500\\
\textcolor{magenta}{ 9}& 3& 2& 18:31:42.420& -1:53:59.847&   2.49 $|$  0.32&   11&  2035.09& 0.669& 0.000\\
\textcolor{magenta}{10}& 3& 2& 18:29:39.387& -1:50:58.225&   1.12 $|$  0.14&   11&  9966.81& 0.766& 2.667\\
	\hline
\textcolor{blue}{11}& 4& 6& 18:31:26.527& -2:7:25.399&   4.74 $|$  0.60&  143&  7304.60& 0.814& 0.083\\
\textcolor{blue}{12}& 4& 6& 18:30:13.549& -2:6:59.828&   4.61 $|$  0.58&  109&  5880.98& 0.440& 0.847\\
\textcolor{blue}{13}& 4& 6& 18:29:10.108& -2:4:47.965&   2.92 $|$  0.37&   45&  6043.60& 0.785& 0.184\\
\textcolor{blue}{14}& 4& 6& 18:31:17.831& -2:15:7.517&   2.37 $|$  0.30&   14&  2846.16& 0.758& 0.000\\
\textcolor{blue}{15}& 4& 6& 18:30:35.244& -2:7:53.587&   1.37 $|$  0.17&   14&  8573.82& 0.678& 0.273\\
\textcolor{blue}{16}& 4& 6& 18:31:56.142& -2:0:34.336&   1.76 $|$  0.22&   12&  4423.46& 0.812& 0.500\\
\textcolor{blue}{17}& 4& 6& 18:29:40.047& -2:7:50.596&   1.96 $|$  0.25&   10&  2988.03& 0.705& 0.000\\
\textcolor{blue}{18}& 4& 6& 18:31:3.668& -2:8:33.347&   1.87 $|$  0.24&   10&  3267.17& 0.775& 0.000\\
\textcolor{blue}{19}& 4& 6& 18:29:55.976& -1:57:48.805&   0.93 $|$  0.12&   10& 13157.63& 0.651& 2.333\\
	\hline
\textcolor{cyan}{20}& 5&11& 18:31:31.056& -2:6:9.457&   2.89 $|$  0.37&   96& 13170.66& 0.785& 0.067\\
\textcolor{cyan}{21}& 5&12& 18:30:26.808& -2:11:5.663&   0.66 $|$  0.08&   14& 36896.97& 0.750& 0.556\\
\textcolor{cyan}{22}& 5&12& 18:30:2.485& -2:3:15.806&   0.69 $|$  0.09&   11& 26457.19& 0.758& 4.500\\
\textcolor{cyan}{23}& 5&12& 18:30:5.246& -2:1:59.236&   0.39 $|$  0.05&   10& 75698.37& 0.705& 0.429\\
	\hline
\end{tabular}}
\end{table*}

\subsection{Grouping of synthetic clusters}\label{chap:mcluster_grouping}

\begin{figure*}
\hspace{-1cm}
\includegraphics[width=0.8\textwidth]{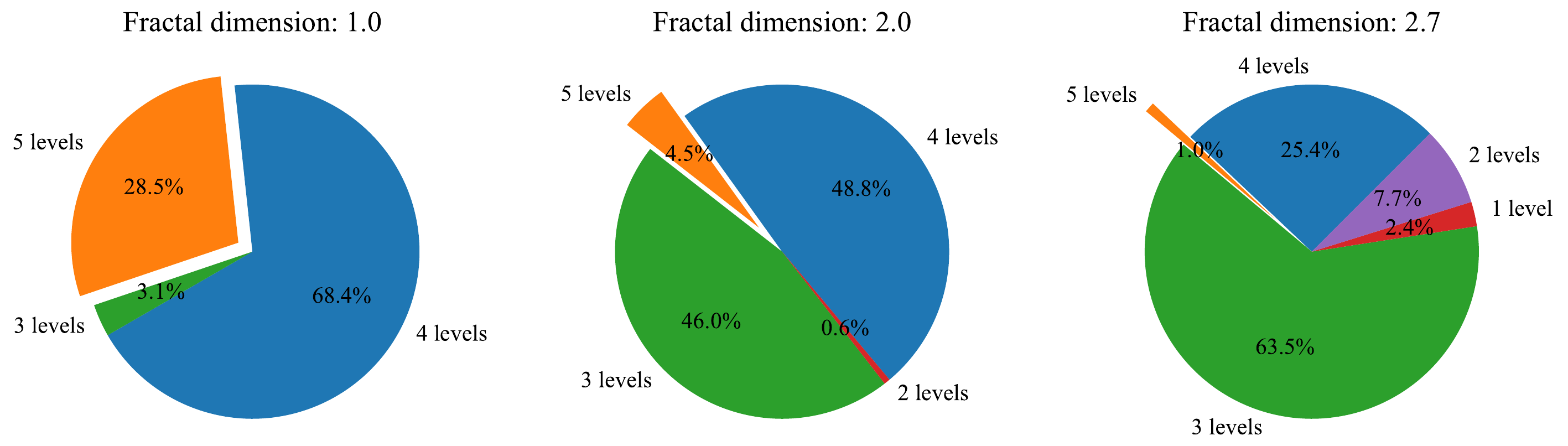}
\caption{
The occupations of different hierarchical levels of the synthetic clusters at the fractal dimension 1.0, 2.0 and 2.7.}
\label{fig:pie}
\end{figure*}

The same grouping method is applied to all the synthetic clusters generated at different fractal dimensions (from 1.0 to 2.0 by a step of 0.1, and 2.7) by \texttt{McLuster}, as described in Sec. \ref{chap:mcluster}.
For each fractal dimension, there are 510 projected 2D distributions, and their resulted ranking of levels may be different.
For the fractal dimensions 1.0, 2.0 and 2.7, we show the occupations of different hierarchical levels of the synthetic clusters in Fig. \ref{fig:pie}.
The percentage of ones containing 5 level structures are 28.5\%, 4.5\%, and 1.0\% for the three fractal dimensions, respectively.
Nevertheless, these statistical results are based on 2D projected maps, which may get some 3D structures degenerated.
For example, at the fractal dimension of 1.0, 145 out of 510 ( 28.5\%) 2D distributions that have 5 levels are from 108 out of 170 ( 63.5\%) 3D synthetic data sets.

Table \ref{tab:grpcounts} lists properties of the synthetic clusters, such as the mean values of hierarchical levels, the numbers of groups and parent groups, and the numbers of groups and parent groups at level 2.
The errors are standard deviations obtained from the 510 projected maps.
The synthetic clusters with smaller fractal dimensions have higher level numbers, and higher numbers of groups and parent groups either in total or just at level 2.
In general, the clusters at fractal dimension 2.0 and 2.7 have more than 3 levels, and the ones at 1.0 have more than 4 levels.
In the cases with higher hierarchical levels, there are more groups identified by our grouping method.
Since nearly all synthetic clusters have at least 2 levels of structures, it is also fair to compare their group numbers at level 2 which reflect the main parts of the cluster.
Coincidentally, the ratio between the number of groups at level 2 and the number of total groups is about 1/4, while the ratio for the parent groups is about 1/2, for each fractal dimension.

\begin{table*}
 	\centering
	 \caption{
	 Statistics of hierarchical structures of both synthetic clusters and the W40 - Serpens South YSO Cluster.
	 N$_\mathrm{groups}$ shows the total number of groups.
	 N$_\mathrm{parent}$ shows the number of groups that have child groups. 
	 L2 denotes properties at level 2. 
	 Slope of $\overline{\Sigma}$ is a fitting result of the mean $\Sigma$ vs. the level numbers, as shown in Fig. \ref{fig:Density}.
	 * About 2.4\% data sets at fractal dimension 2.7 have only 1 hierarchical level, and thus their slopes of $\overline{\Sigma}$ are set to be 0 during the calculation.
	 }
	 \label{tab:grpcounts}
	\hspace*{-2cm}\begin{tabular}{c|ccc|cc|c}
	\hline
Sample& Levels & N$_\mathrm{groups}$ & N$_\mathrm{parent}$ & L2 N$_\mathrm{groups}$ & L2 N$_\mathrm{parent}$ & Slope $\overline{\Sigma}$\\
	\hline
All Fractal 1.0 maps&   4.3 ± 0.5& 43 ± 4& 13.6 ± 2.6& 12 ± 5&  7.6 ± 2.4&   0.68 ± 0.11\\
All Fractal 2.0 maps &  3.6 ± 0.6&  21 ± 4&  3.5 ± 1.2&   5 ± 3&  1.7 ± 0.8&   0.21 ± 0.05\\
All Fractal 2.7maps &  3.2 ± 0.7& 15 ± 5&  2.3 ± 0.7&   3 ± 2&   1.0 ± 0.3&  *0.10 ± 0.04\\
	\hline
 A typical 5-level F1.0 (See Fig. \ref{fig:Mcluster_Grouping}) &     5&    49&    17&    11&     7&  0.57 \\
 A typical 5-level F2.0 (See Fig. \ref{fig:Mcluster_Grouping}) &     5&    23&     6&     2&     1&  0.17 \\
 A typical 5-level F2.7 (See Fig. \ref{fig:Mcluster_Grouping}) &     5&    18&     4&     3&     1&  0.07 \\
	\hline
W40 - Serpens South YSO Cluster &     5&    23&     5&     4&     1&  0.41 \\
	\hline
\end{tabular}
\end{table*}

\begin{figure*}
\hspace{-0.5cm}
\includegraphics[width=0.95\textwidth]{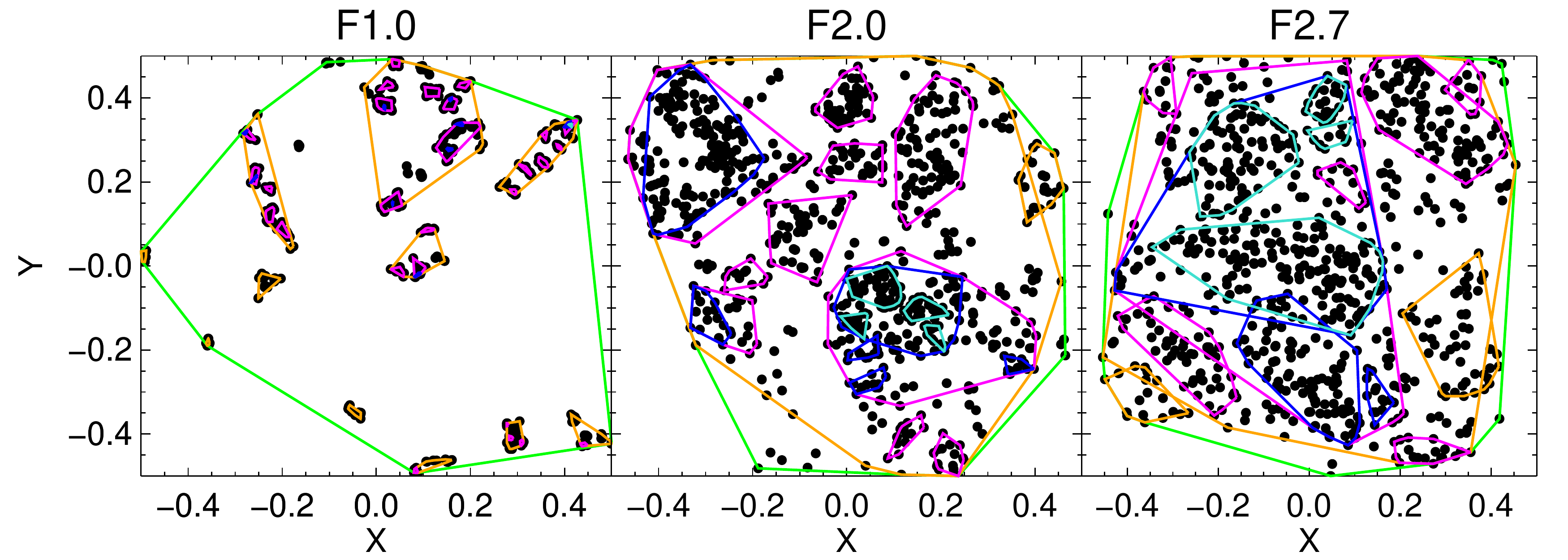}
\caption{
Three selected fractal models of clusters with dimension 1.0, 2.0, and 2.7. 
Colors of levels are the same as in Fig. \ref{fig:YSOs_Grouping}.}
\label{fig:Mcluster_Grouping}
\end{figure*}

Nevertheless, since the observed W40 - Serpens South YSO Cluster shows 5 levels, we select the synthetic clusters with 5 levels for comparison.
Three representative synthetic clusters with fractal dimensions of 1.0 (labeled as ``F1.0'' hereafter), 2.0 (``F2.0'') and 2.7 (``F2.7'') are explored.
Their hierarchical structures are shown in Fig. \ref{fig:Mcluster_Grouping}, and the parameters of levels and group numbers are also listed in Table \ref{tab:grpcounts}. 

For the F1.0 model, the groups are sparse but distinctive, at least for the levels 1--3.
The groups at the levels 4--5 have very small sizes and can hardly be identified through visual check.
The mean critical edge lengths for different levels in F1.0 are 263.2$''$, 50.8$''$, 20.9$''$,  and 15.1$''$, which has a degressive factor of $\sim$3.0.

For the F2.0 model, the groups are not as compact as those in F1.0, but still distinctive, especially for the level 3.
The groups at the levels 4--5 have relatively larger sizes.
The mean critical edge lengths in F2.0 are 204.3$''$, 121.3$''$, 98.7$''$, and 69.9$''$, with a degressive factor of $\sim$1.4.

For the F2.7 model, the points nearly suffuse the entire region, and the groups can not be easily separated through visual check.
The mean critical edge lengths in F2.7 are 161.5$''$, 129.7$''$, 119.2$''$,  and 107.2$''$, with a degressive factor of $\sim$1.1.
The critical lengths do not change noticeably, and the sizes of groups in the level 4 and 5 are also large.

In general for three representative models, the numbers of their groups or parent groups are similar to the mean values of clusters at corresponding fractal dimensions (Table \ref{tab:grpcounts}), even though we deliberately chose examples with five distinguishable hierarchical levels.
In comparison with the above models, we can see that the degressive factor of critical length of YSO cluster is $\sim$1.8, which lies between F1.0 and F2.0 models.
Besides this, the structural characteristics of the YSO cluster in the  W40 - Serpens South region are similar to the F2.0, or close to the mean parameters of all the synthetic clusters at the fractal dimension 2.0.

\section{Discussions about fractal dimensions}\label{sec:Fd}

\subsection{$Q$ parameter}\label{chap:qparameter}

\begin{figure*}
\centering
\includegraphics[width=0.95\textwidth]{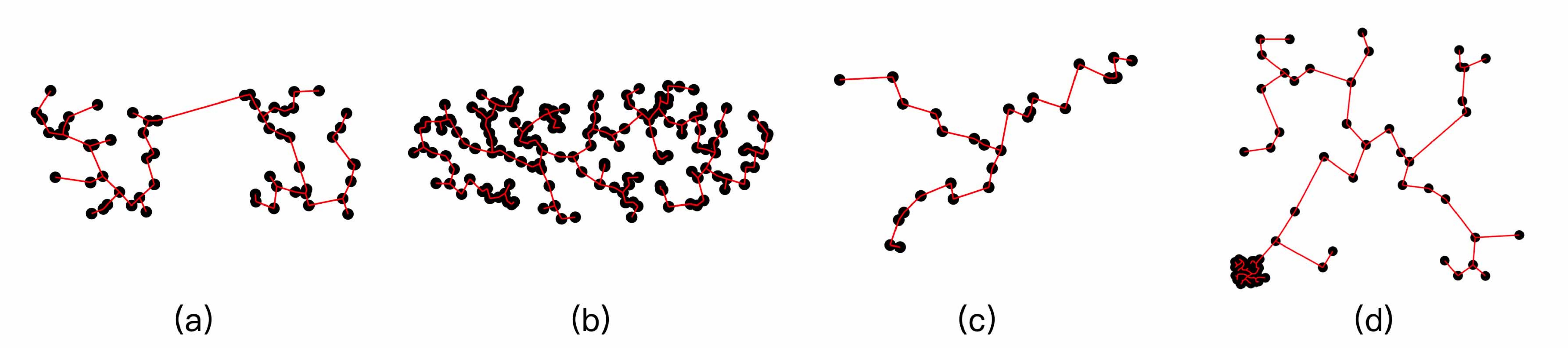}
\caption{
Four cluster patterns with low $Q$ values ($<$0.6) yet are not generated by fractal models. 
Total member points are 60 (30+30), 160, 28, and 101 (core 40+external 60) for (a) a pair of cluster cores, (b) elongated cluster, (c) cluster with split branches, and (d) cluster with an outward over-dense region, respectively. }
\label{fig:q_example}
\end{figure*}

In this section, we calculate $Q$ parameter of the observed and synthetic data.  
This parameter is commonly used to describe the structure of a cluster, and can tell apart those of centrally condensed or fractal distributed structures.
The $Q$ parameter is defined as \citep{Cartwright2004}
\begin{equation}
Q = \overline{m}/\overline{s},
\end{equation}
where the $\overline{m}$ is a normalized mean value of all the MST edge lengthes, and the $\overline{s}$ is a normalized mean value of the separated distances of all the member points.
When $Q$ is larger than 1, the point distribution is more concentrated in the center. 
For example, the  `Example 1' in Section \ref{chap:GroupingMethod} has an overall $Q$ value of 1.215.
When $Q$ is smaller than 0.6, the distribution is supposed to have more substructures.
The `Example 2' has a low $Q$ value of 0.442.
When $Q$ is in-between, the distribution has a more flat or uniform structure.
Fig. \ref{fig:q_example} shows four illustration patterns with low $Q$ values ($<$0.6), in which the second elongated pattern has been thoroughly studied by \cite{Bastian2009}.

We further check the capability of the $Q$ parameter based on synthetic clusters with different $N$ members (from 10 to 800) and different fractal dimensions (from 1.0 to 3.0).
When $N>50$, the $Q$ shows a positive correlation with the fractal dimensions.
Nevertheless, the dispersion of the correlation is significant, and the difference between fractal dimension 2.0 and 2.7 can be hardly distinguished.
The fractal dimension 1.9 is roughly corresponding to the $Q=0.6$, which is also consistent with the notional relationship of dimension and $Q$ derived in \citep{Cartwright2004}.
The points with $N<50$ can hardly express sub-structures, and thus the $Q$ values are generally larger than 0.6.

\begin{figure*}
\centering
\includegraphics[width=0.95\textwidth]{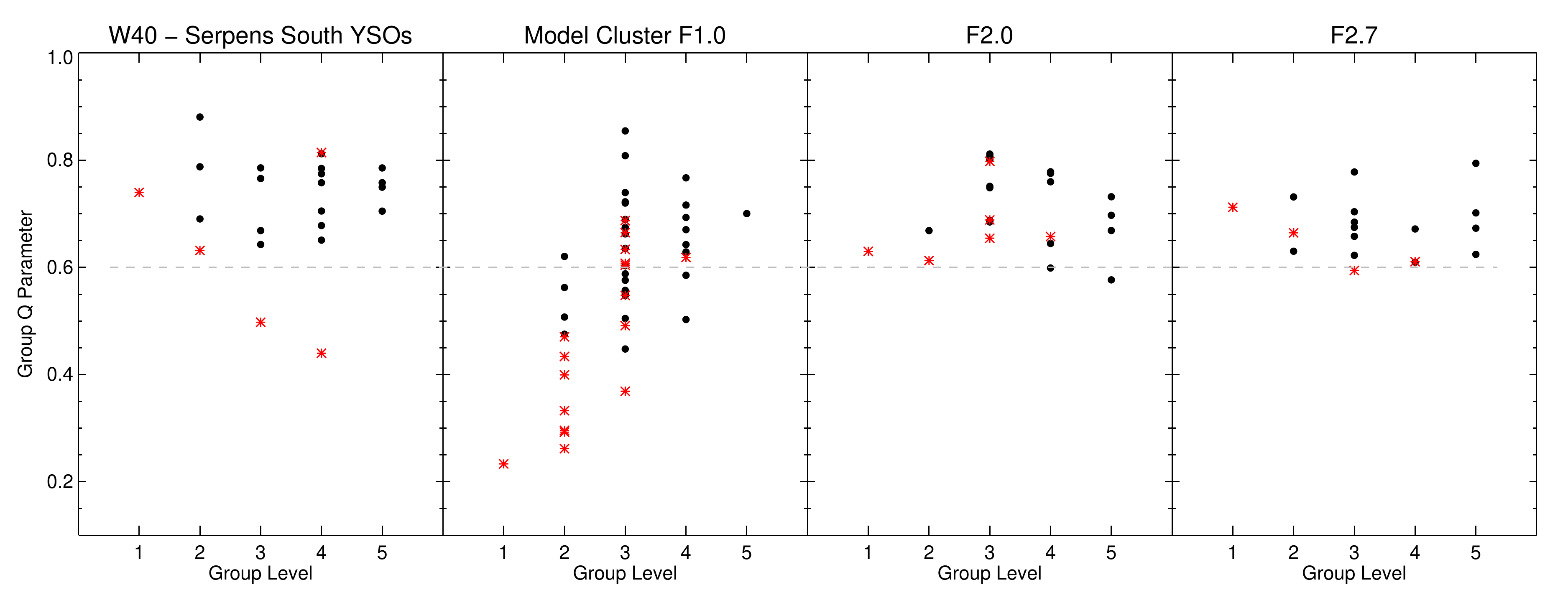}
\caption{
$Q$ parameter as a function of the group level. 
Red asterisks indicate the parent groups, and black dots are the child groups. 
The dash line shows $Q=0.6$.}
\label{fig:Q_param}
\end{figure*}

Fig. \ref{fig:Q_param} shows all the $Q$ values of groups at different levels in the W40 - Serpens South region, which are also listed in Table \ref{tab:grpcounts}.
In general, the $Q$ values are more meaningful for the groups 1, 2, 6, 11, 12, and 20 that have $N>50$. 
At the level 1, the $Q$ of group 1 is about 0.75, a relatively high value, suggesting that the whole region has a flat geometry on average.
At the levels 2--3, the $Q$ values of the parent groups 2 and 6 are smaller, indicating more complex sub-structures.
At the level 4, the $Q$ values of the groups 11 and 12 are well separated, and thus their fractal dimensions are distinguishable.
The group 11 has only one child group 20 at the level 5, while the group 12 has three cluster cores distributed at two ends of its elongated shape.
As a result, the former has a higher $Q$, and the latter has a lowest value among all groups.
The $Q$ values of all other child groups with $N<50$ are roughly between 0.6 and 0.8, indicating a more flat or random distribution.

Fig. \ref{fig:Q_param} also shows the $Q$ values of all groups in the F1.0, F2.0, and F2.7 cases.
Since the fractal model does not produce central condensed distributions, the $Q$ values should be less than 1.
In both the cases of F2.0 and F2.7, the $Q$ values are roughly between 0.6 and 0.8, but the difference between them is not significant.
In the F1.0 case, many groups have lower $Q$ values, especially at the levels 1--3 where the structures are filamentary (Fig. \ref{fig:Mcluster_Grouping}) and many groups have $N>50$.
Generally the $Q$ values increase with the hierarchical levels, though at each level the fractal dimension is similar $\sim$1.0.
This may be caused by our grouping method, since it alway cuts down stretched branches and is likely to obtain a higher $Q$.

The trend of $Q$ values of the observed YSO groups in W40 - Serpens South is more consistent with that of F2.0 or F2.7.
However, the YSO group 6 which represents for the clusters in the W40 and the Serpens South regions, and the group 12 that for the clusters in the Serpens South region specifically, have lower $Q$ values, suggesting additional substructures.
Therefore, the pattern of the observed YSO cluster W40 - Serpens South in is not likely described by a single fractal dimension.

\subsection{Profiles of surface densities ($\Sigma$) }\label{chap:surfacedensity}

\begin{figure*}
\centering
\includegraphics[scale=0.65]{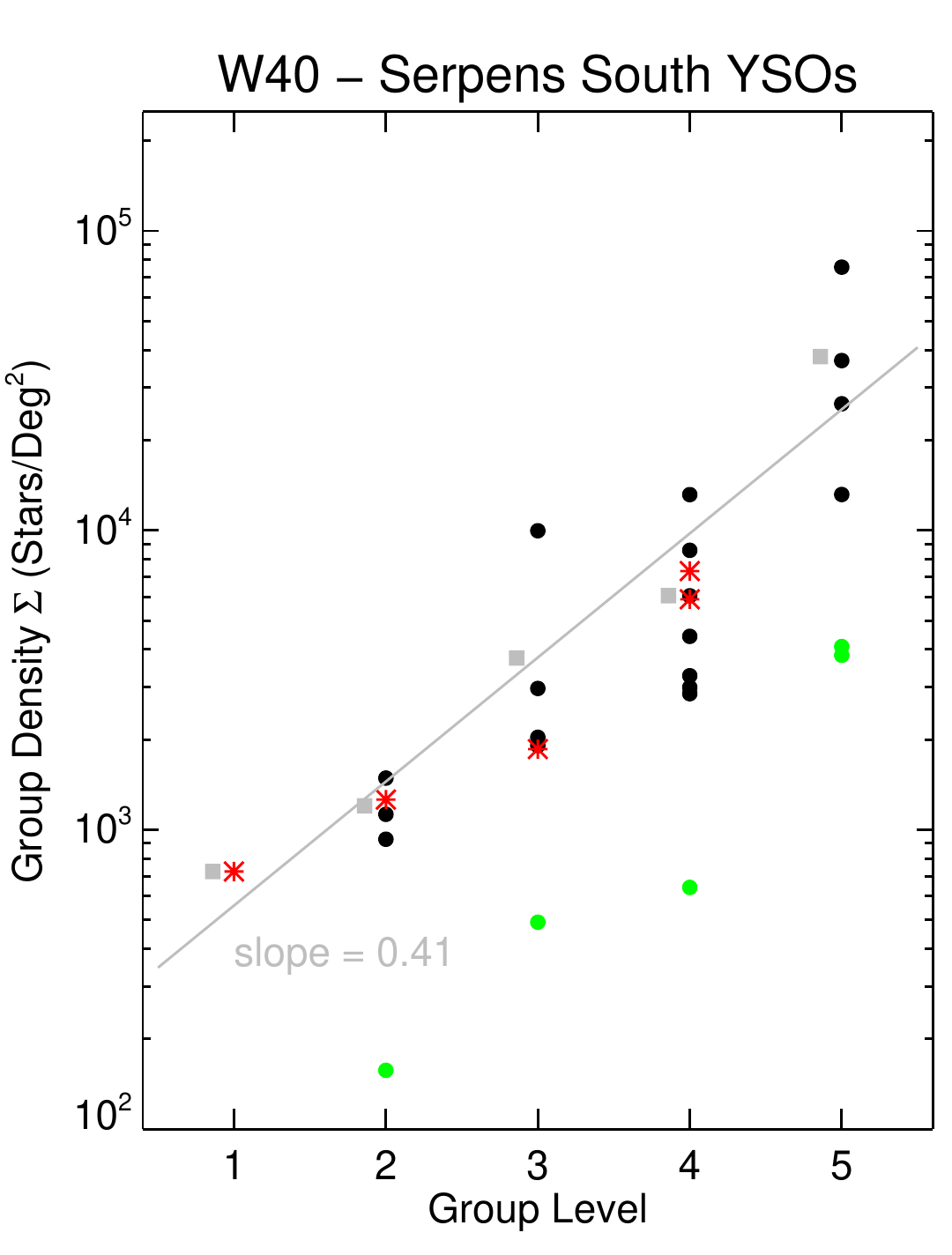}
\includegraphics[scale=0.447]{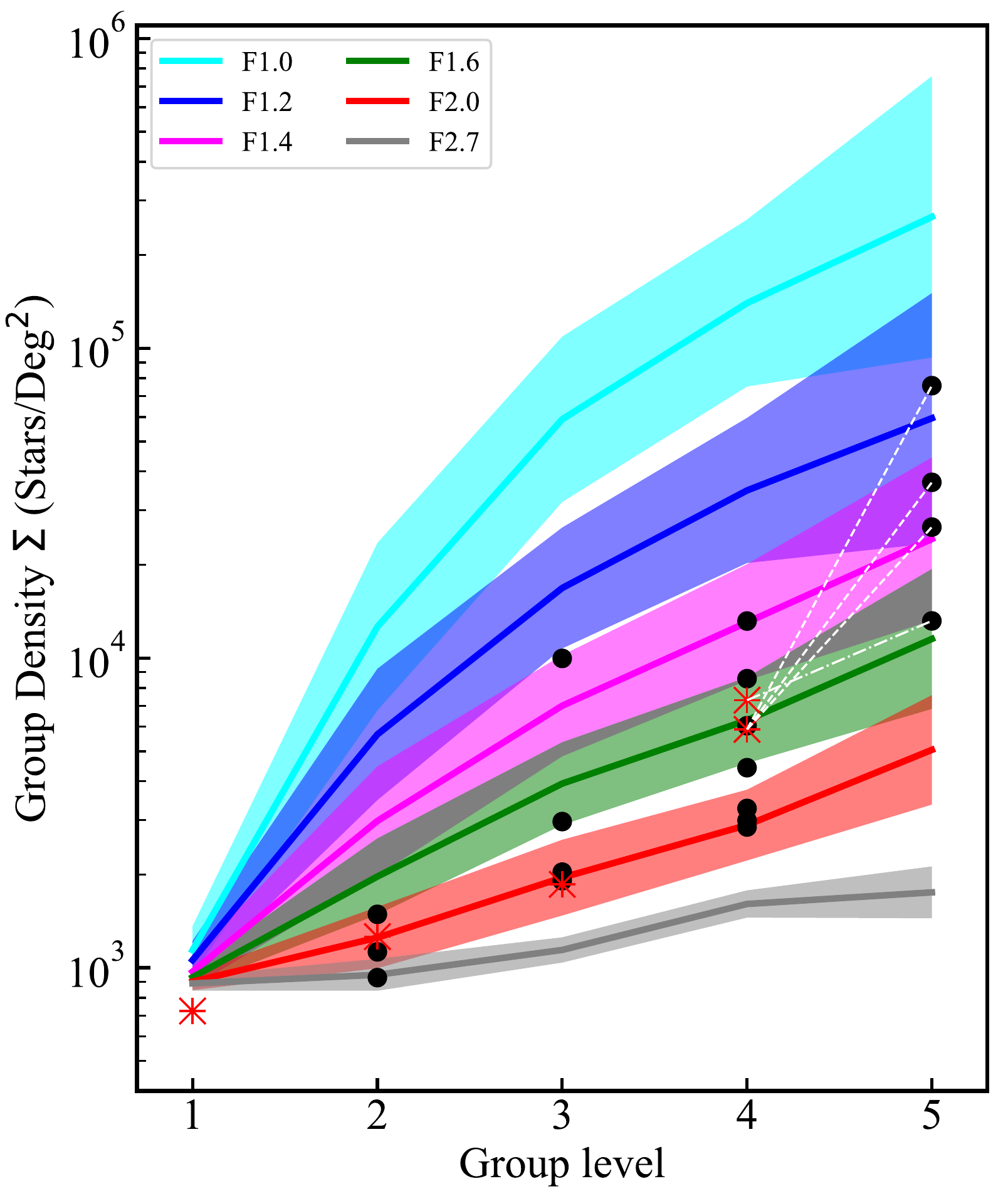}
\includegraphics[scale=0.8]{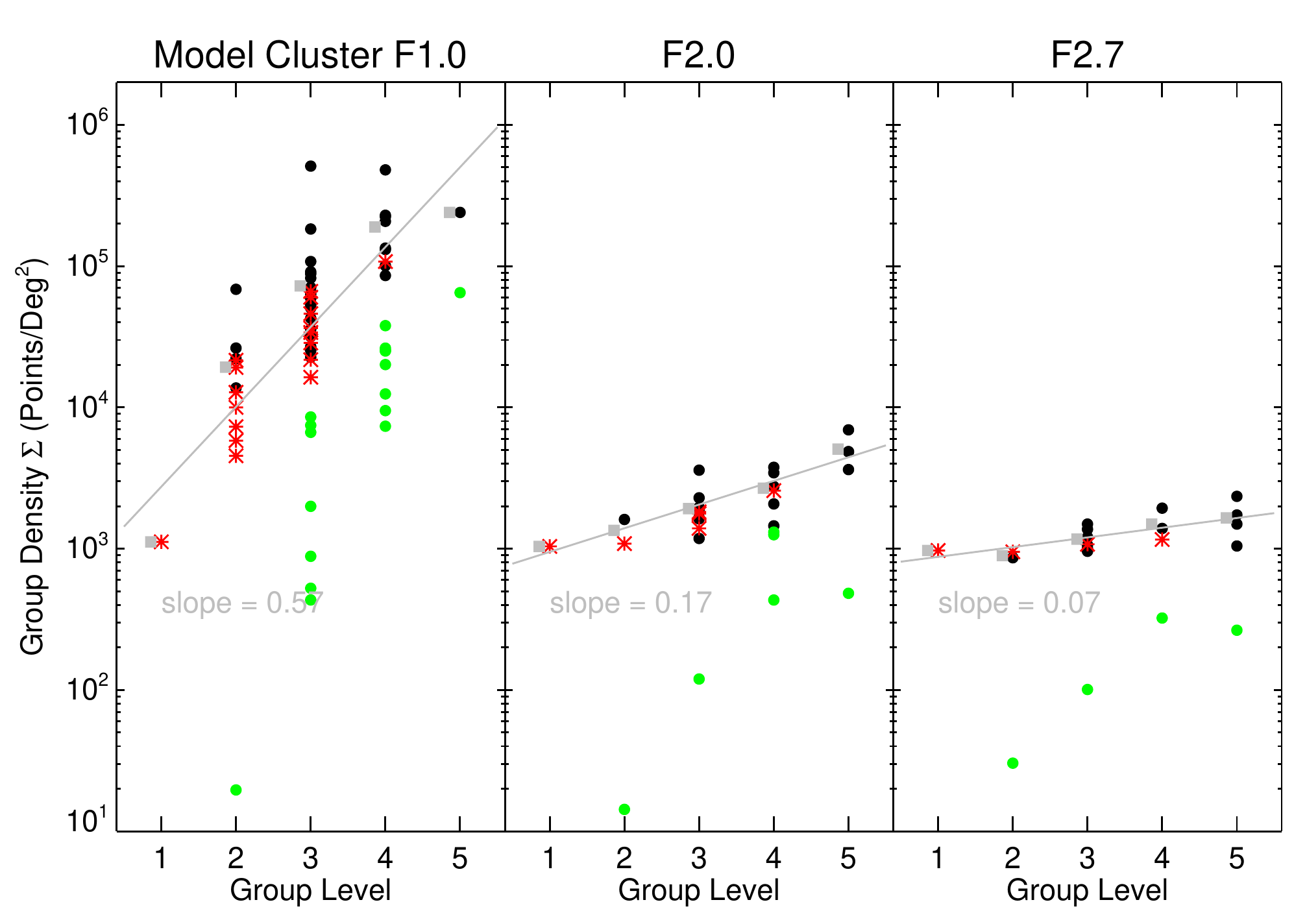}
\caption{ Diagrams of surface density $\Sigma$ versus group levels. 
The upper panel is for the YSOs in the  W40 - Serpens South region, and the lower panel is for the three synthetic fractal clusters F1.0, F2.0, and F2.7.
Red asterisks,  black dots, and green points are representing the parent groups, the child groups, and the ``leftover'' points, respectively.
Grey squares are the mean $\Sigma$ values of groups at each level, and the grey line is the best-fit line with the slope marked.
In the upper right plot, the $\overline{\Sigma}$ profiles of synthetic clusters at different fractal dimensions, together with their derived standard deviations are illustrated by color stripes for the comparison.
The group 11 and its child group 20 are linked by a white dash-dotted line, which are from the W40 region, while the group 12 and its three child groups 21--23 are linked by white dash lines, which are from the Serpens South region.}
\label{fig:Density}
\end{figure*}

The groups at higher levels are supposed to have higher surface densities ($\Sigma= \frac{N}{A}$ in Eqn. \ref{eq:cdf}).
We calculate the $\Sigma$ values of all YSO groups in the W40 - Serpens South region, through dividing their $N$ by the included area $A$ inside their convex hull.
The results are listed in the third-to-last column in Table \ref{tab:grpinfo}. 
The $\Sigma$ values of the leftover YSOs that are not classified as members in child groups are calculated in a similar way, but using the leftover area between the parent group and its child groups.
Fig. \ref{fig:Density} shows the $\Sigma$ versus the hierarchical levels. 
We note that at the bottom of the convex hull of the group 1, some vain area that is not observed has been included.
Thus its $\Sigma$ decreases by a few percent.

There is a clear trend that the $\Sigma$ of either parent or child groups are getting higher at higher levels, as well as the leftover sparse YSOs .
This indeed suggests a hierarchical structure of the YSO distribution in the entire W40 - Serpens South region.
We obtain a slope of 0.41 by fitting linearly the mean values of the $\Sigma$ of all groups.
The scattering at levels 3--5, however, is still as large as one order of magnitude.
At level 3, the highest one is from group 10, which is a part of the Serpens South region.
At level 5, the three highest points are all from the Serpens South region, while the lowest one is from the W40 region.

Then, we compare the $\Sigma$ values from the observed YSO cluster in W40 - Serpens South with $\Sigma$ profiles at different fractal dimensions, which are derived from all synthetic clusters with 5 levels.
These synthetic data sets can be well separated, suggesting the $\Sigma$ profiles can be an excellent tracer for the fractal dimension.
At the levels 1--3, the YSO densities are close to the mean profile at the fractal dimension 2.0, except for a higher $\Sigma$ point of the group 10 at the level 3.
Since the group 1 occupies some vain area, its $\Sigma$ at the level 1 is slightly smaller than the values in those $\Sigma$ profiles of fractal clusters.
At the levels 4--5, the YSO densities increase quickly, matching the density profiles of models at the fractal dimensions 1.6 and 1.4. 
If we treat the W40 and the Serpens South regions separately, the former is close to the fractal 1.6 model and the latter is close to the fractal 1.4 model.
The three higher points at the level 5 (the groups 21--23), as well as the highest point at the level 3 (the group 10), all belong to the Serpens South region.
In general, the fractal dimension is larger in the periphery region, but goes smaller around the core regions.
The YSO distribution needs to be described by patterns from multi-fractal dimensions.

Fig. \ref{fig:Density} also shows the $\Sigma$ values of groups in F1.0, F2.0, and F2.7.
They are generally following the $\Sigma$ profiles of synthetic clusters at the same fractal dimension, but in the figure the scattering of $\Sigma$ values of the parent and child groups, and even the leftovers can be better seen.
The $\Sigma$ values of the child groups are generally larger than that of the parent groups.
By fitting the mean value at each level, we obtain the slopes of F1.0, F2.0, and F2.7 are 0.57, 0.17, and 0.07, respectively (Table \ref{tab:grpcounts}).
The slope of the YSO cluster is 0.41, which is somewhere between the slopes of F1.0 and F2.0.
We also check the mean profile slopes of all synthetic clusters at the fractal dimensions 1.0, 2.0, and 2.7,  and the resulting slopes are slightly steeper than that of F1.0, F2.0, and F2.7 (Table \ref{tab:grpcounts}).

\subsection{Correlations with the YSO classification}

The above discussion on grouping and estimations of the hierarchical structures of YSOs haven't been correlated with the YSO classification yet.
In this section, we discuss the correlation between the fractal pattern around different structures and the age of YSOs.
The ratios of Class I (including deeply embedded sources) and Class II (including transition disk sources) of all YSO groups are listed in the last column of Table \ref{tab:grpinfo}, which can be used to infer the YSO age.

The groups 10, 12, and 19 at the levels 3--4 have the highest ratio values, if the groups 21--23 at the level 5 are not considered since they are all from the parent group 19.
It means that these groups at different levels should have the youngest YSOs. 
They are actually different parts of the Serpens South region (Figure \ref{fig:YSOsMST_demo}).
They may form simultaneously, but do not have to be physically correlated with each other.
As a result, with class ratio of the group 12 to be 0.847, the Serpens South region rank among the youngest regions \citep{Gutermuth2008serpens, Li2019}. 
It also has the lowest fractal dimension ($\sim$1.4) either by the $Q$ values or the $\Sigma$ profile.
The W40 region has the group 11 and its child group 20, which has class ratios $<$ 0.1, suggesting the age of about several Myr (Sun et al. 2022, submitted to MNRAS).
The fractal dimension in the W40 region is about 1.6 by the $\Sigma$ profile, and the fractal dimension is larger than that in the Serpens South region.

Except for the groups in the Serpens South region, the groups 3, 8, and 16 also have the Class I/II ratios larger than 0.5, suggesting that they are new born clusters too.
The group 3 is at level 2, but does reside in a relatively isolated dense cloud.
The group 8 is inside the Donut cloud, and the group 16 resides clearly on a filamentary cloud.
They are just small ($N$) child groups at different levels, and do not have any sign of higher fractal dimension.
Like group 3, 8, and 16, group 13 also has some correlated gas left, but is a more evolved and independent collection .

The groups 4, 7, 9, 14, 17, and 18 do not contain any Class I YSO. 
These clusters are more evolved, and their local molecular gas is likely exhausted or dispersed.
As said, there are no reliable tracers of fractal dimensions for these individual small child groups.
In contrast, there are some `leftover' protostars in the periphery regions, though not belonging to any groups.

We notice that the size of the group 20, which is $\sim$0.37 pc, is 5 times of the mean size of $\sim$0.07 pc of the groups 21--23.
The velocity dispersion of prestellar cores in a small region (e.g., $<$1 pc) is reported to be $\sim$0.5 km/s in Taurus \citep{Qian2012}.
That is $\sim$0.63 pc distance variance for 5 Myr, suggesting that W40 cluster could have complex structures and more sub-clusters $\sim$5 Myr ago, just like in the Serpens South cloud.

\section{Summary}\label{sec:sum}
Young stellar clusters are believed to inherit the hierarchical structures of their natal molecular clouds, but such structures would not last for long.
This work investigates the spatial distribution of protostars and pre-main sequence stars within one molecular cloud, which is the intermediate link revealing how the hierarchical structures evolve from the natal clouds to the stellar clusters.
The W40 - Serpens South Region ($\sim7\times9$ pc$^2$) of the Aquila Rift is selected as our testbed, and 832 young stellar objects (YSOs) have been identified in this region.
The data has advantages in both sample number and the slight age difference of the two sub-regions.
Mainly three dimensions of fractal model, generated by the \texttt{McLuster}, are used to characterize hierarchical structures of this young cluster and check whether a correlated dimension exists.
The main results are summarized as follows:

1. We demonstrate a means of extracting hierarchical structure from a 2D point distribution algorithmically.
We use a MST-based method to group stars into several levels by continuously cutting down edges (i.e., distance between to nearest stars) that are longer than a critical value, until the distribution of each extracted star group seems random or the number $N<20$.
This grouping method determines the hierarchical levels of the point distribution, and recognizes those concentrated structures as groups at each level.

2. According to this grouping method, the 832 YSOs are divided into five levels with 23 groups, in which the groups representing core clusters in specific the W40 region and the Serpens South region are recognized and separated at the level 4.
While the W40 core group has one child group at the level 5, the Serpens South filamentary group has three child groups, indicating a more complex structure.

3. In addition, we generate thousands of synthetic data sets at various fractal dimensions, and apply the same grouping method to explore their hierarchical structures.
The synthetic clusters with lower fractal dimensions have a higher percentage of 5-level hierarchical structures and more groups and parent groups.
The lower fractal dimension clusters tend to have lower $Q$ at the first parent group, and higher $Q$ with groups in subsequent levels, till groups with number $N<50$ roughly have undistinguishable $Q$ values between 0.6--0.8.
The surface density $\Sigma$ profiles of the synthetic clusters have an apparent correlation with the fractal dimensions.
When the fractal dimension is smaller, the profile tends to be higher and have a steeper slope versus the group levels, vice versa.

4. By comparing the $Q$ parameter and the $\Sigma$ profiles of the observed and the synthetic data, we find that the YSO observation matches synthetical models with multi fractal dimensions.
In the periphery region where the molecular clouds are relatively diffuse, the YSO structure is close to that of the synthetic data at a fractal dimension of 2.0, while in the core regions the fractal dimension is clearly lower.
In the Serpens South region where the YSOs are younger than 1 Myr, the fractal dimension is close to 1.4, and in the W40 region where the YSOs have evolved for a few Myr, the fractal dimension is close to 1.6.

Therefore, the YSO clusters may inherit the fractal pattern of the dense part of molecular clouds, but such pattern dissipates slowly in several Myr. 

\section*{Acknowledgements}
We thank the anonymous referee for giving us a thorough reading and precious comments, which improve and make the manuscript more complete.
J. Sun acknowledges financial support from the China Scholarship Council (CSC), and is also supported by the NSFC of China grant No.12073079 and No.11973090.
H. Wang acknowledges the support by NSFC grants No.11973091.
S.N. Zhang acknowledges the support from NSFC grant No.11573070.
R.A. Gutermuth gratefully acknowledges funding support for this work from NASA ADAP award NNX17AF24G.
This research has made use of data from the Herschel Gould Belt survey (HGBS) project (http:// gouldbelt-herschel.cea.fr).


\bibliographystyle{mnras}
\bibliography{MNscript_cluster_structure} 








\label{lastpage}
\end{document}